\documentstyle[12pt]{article}

\newcommand \be {\begin{equation}}
\newcommand \ee {\end{equation}}
\include{dspace12}
\include{epsf}
\begin{document}
\begin{center}
{\bf{\large A  Connection of Apparent Horizon and Naked Singularities}}
\end{center}

\begin{center}
{Sukratu  Barve}\footnote{sukratu@imsc.ernet.in}\\
  The Institute  of Mathematical Sciences\\
   Chennai\\
\end{center}

\abstract
          {We show  that  the behaviour of  the outgoing radial null
 geodesic congruence on the apparent horizon is  related to the  property
 of nakedness in spherical dust collapse justifying the  difference in the 
Penrose diagrams in the naked and covered dust collapse scenarios. 
 We provide arguments  suggesting that the relationship could be generally 
valid.}
\newpage
\section{Introduction}

        Consider a cloud of matter (regular initial Cauchy data) collapsing indefinitely under its own gravity. A singularity eventually develops in the spacetime and it is indicated by the divergence of the Kretschmann scalar. In the advanced stages of collapse trapped regions are formed \cite{schoen}, \cite{penrose}
 and there exists a null ray which marginally escapes to infinity (event horizon).
 It is not clear whether the singular boundary is entirely surrounded by the trapped region. In other words, it is not known if a portion of the boundary is exposed in the untrapped region and non-spacelike geodesics can emanate from it (naked singularities).
        
 Exact solutions to Einstein's field equations with certain kind of source 
terms are known to exhibit both naked and covered singularities depending upon
 the sort of regular initial data chosen. Little progress has been made 
regarding a general classification of initial data according to the covered or
 naked 
 consequence of the evolution. It is not known if the data leading to naked 
singularities has zero measure in the set of all possible (or all possible
 physically relevant) Cauchy data. 

 The complexity of the problem lies in the fact that the Cauchy initial
 value problem for the Einstein field equations with sources is less
 tractable. The systematics available about the general problem is far too
 less for any implication for questions like formation of naked singularities.
  For instance,
 even the well posedness of the problem is not self evident and has to
 be proved independently for different types of sources \cite{waldgr}, \cite{haw}, \cite{gara}. It would be indeed difficult to find or even expect
 a conserved or monotonically behaving function of the Cauchy surface
 with respect to its evolution, which could be expected to provide
 insight into the process of creation of naked singularities.

   As a result of this difficulty, a large number of investigations that have been carried out have been 
 concerning  certain exact solutions or numerical simulations. Issues like
  strength of the singularities, genericity, behaviour
 with respect to change of source etc. have been studied in examples
 like dust, null dust, perfect and imperfect fluids and scalar fields.
 However, they do not suggest any 
 typical geometrical feature which could be expected to arise before
 a naked singularity forms ( In case of a singularity the singularity theorems
  make use of a typical geometrical feature viz. trapped regions to prove its existence). Such a feature indicating the presence of a naked singularity would be interesting in the light of the 
 Hoop conjecture or the issue of isoperimetric inequalities and could lead to
 some geometrical insight into the process. 
 The lack of indication of existence of such a feature
 is evident in the fact that one is forced to check for the existence of naked
 singularities
 in a direct manner even in particular examples. To be precise, one checks if
non space-like geodesics emerge from the singular boundary using differential
 geometry and the form of the metric in the example.  

This paper is a first step towards an indirect criterion. Preferably, the criterion should be applicable away from the singular boundary. That is a non-local
 problem and given the difficulties of Cauchy evolution, there is no indication available for what the criterion could be. It would be perhaps appropriate in such a situation to examine regions near the singularity. There are also parts of the 
 singularity from
 which geodesics cannot escape. Some criterion applicable to such a portion
 would also be significant. It would indicate that the information about
 the exposure of a part of the boundary is contained elsewhere on the boundary.

We illustrate the main features of spherical dust collapse in radial co-ordinates (figures 1 and 2) and in causally correct Penrose diagrams (figures 3 and 4).

\begin{figure}[h]
\parbox[b]{20.99cm}
{
\epsfxsize=12.95cm
\epsfbox{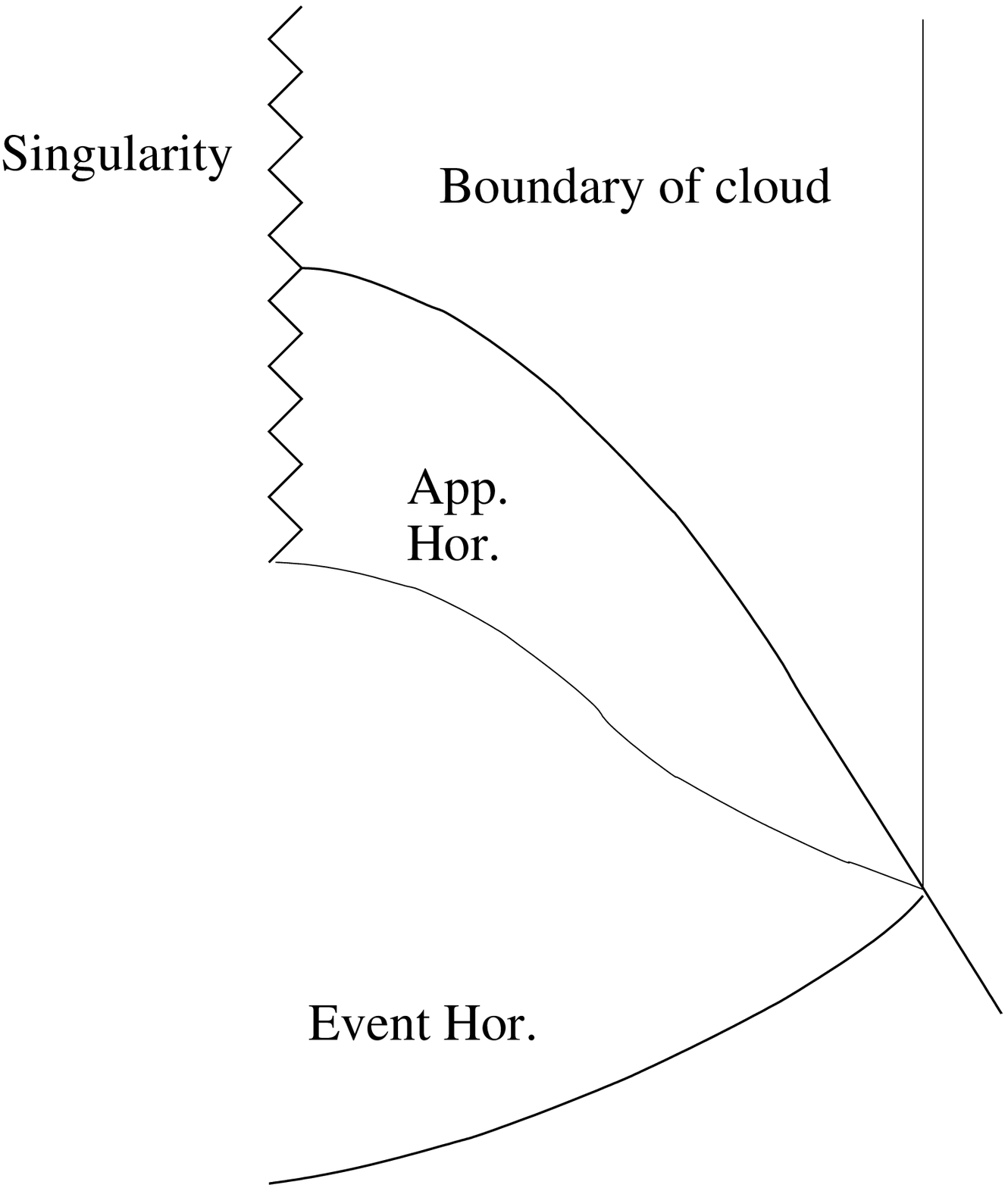}
}
\caption{Collapse of spherical dust leading to a covered singularity}
\end{figure}

\begin{figure}[h]
\parbox[b]{20.99cm}
{
\epsfxsize=12.95cm
\epsfbox{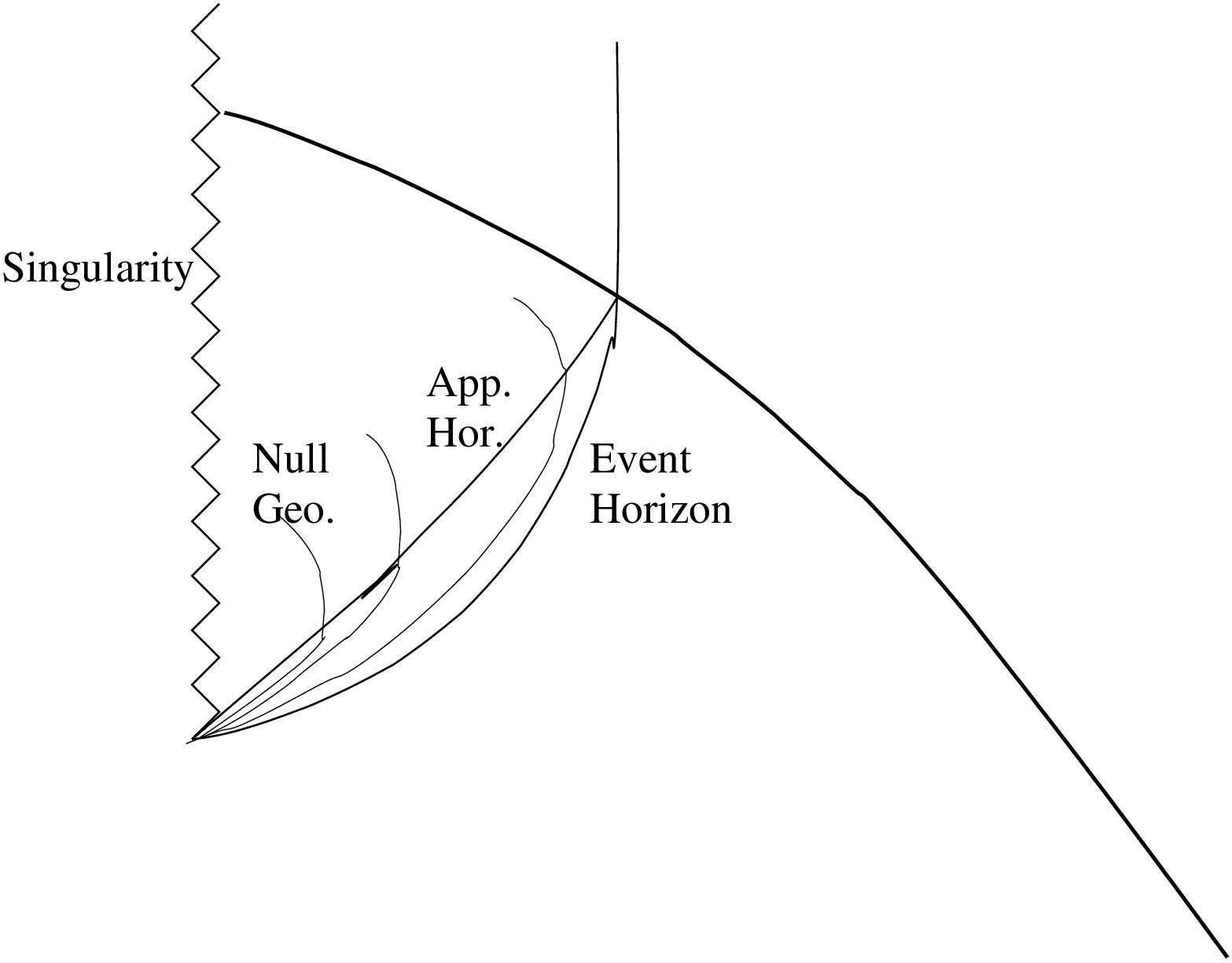}
}

\caption{Collapse of spherical dust leading to a naked singularity}
\end{figure}

\begin{figure}[h]
\parbox[b]{20.99cm}
{
\epsfxsize=12.95cm
\epsfbox{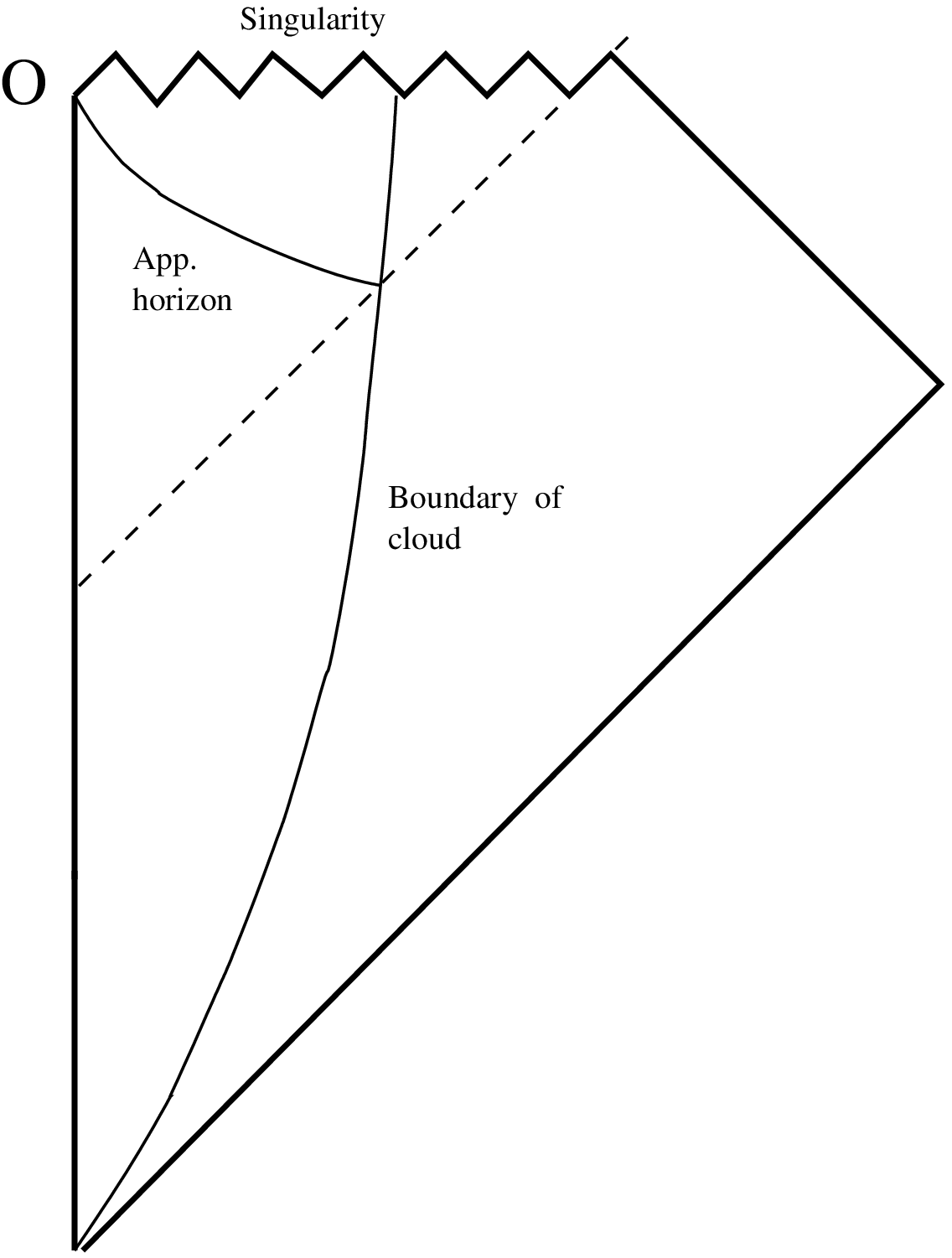}
}
\caption{Penrose Carter diagram for collapse of spherical dust leading to a 
covered singularity}
\end{figure}

\begin{figure}[h]
\parbox[b]{20.99cm}
{
\epsfxsize=12.95cm
\epsfbox{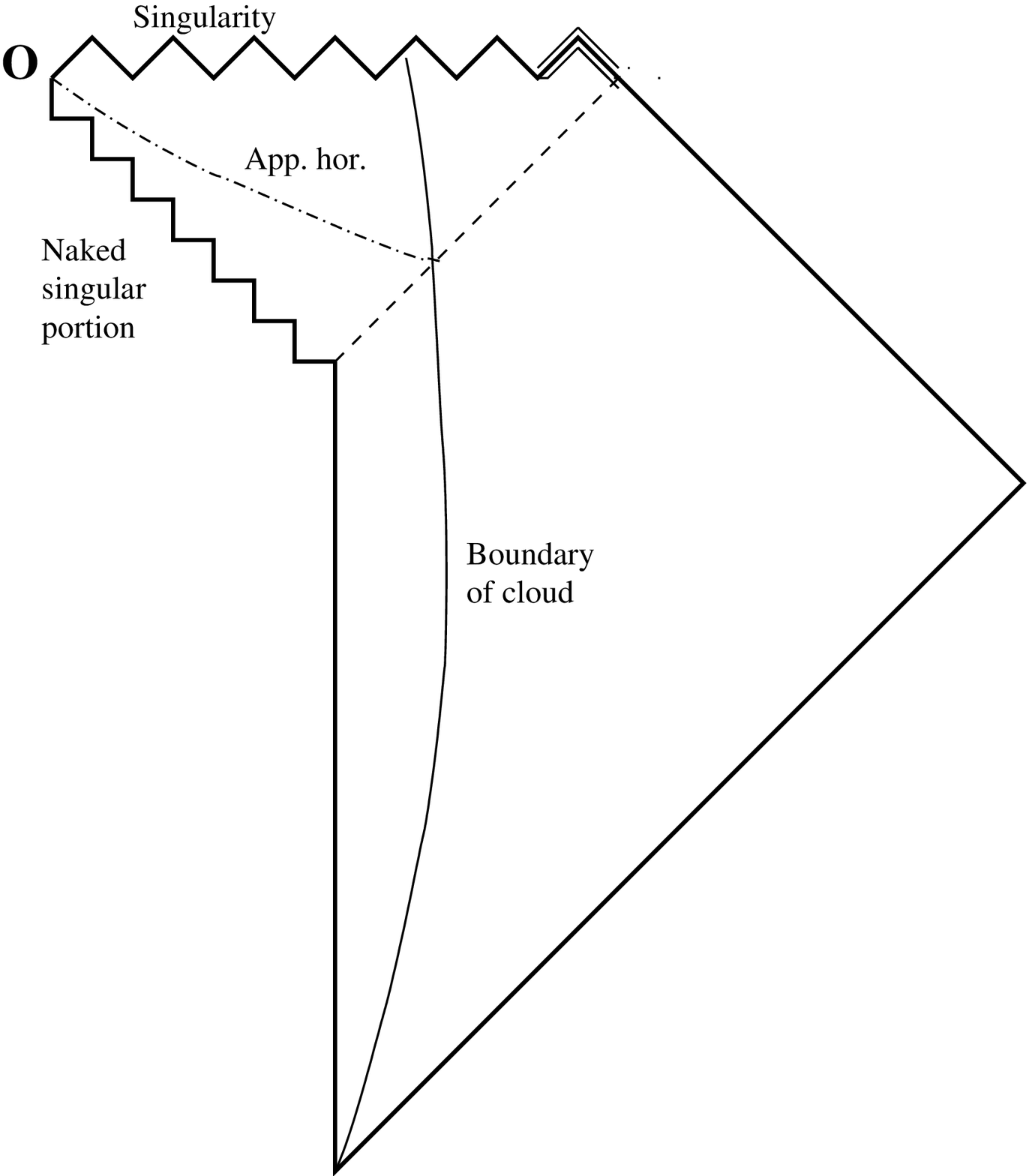}
}
\caption{Penrose-Carter diagram for collapse of spherical dust leading to a 
(locally) naked singularity}
\end{figure}

\begin{figure}[h]
\parbox[b]{20.99cm}
{
\epsfxsize=12.95cm
\epsfbox{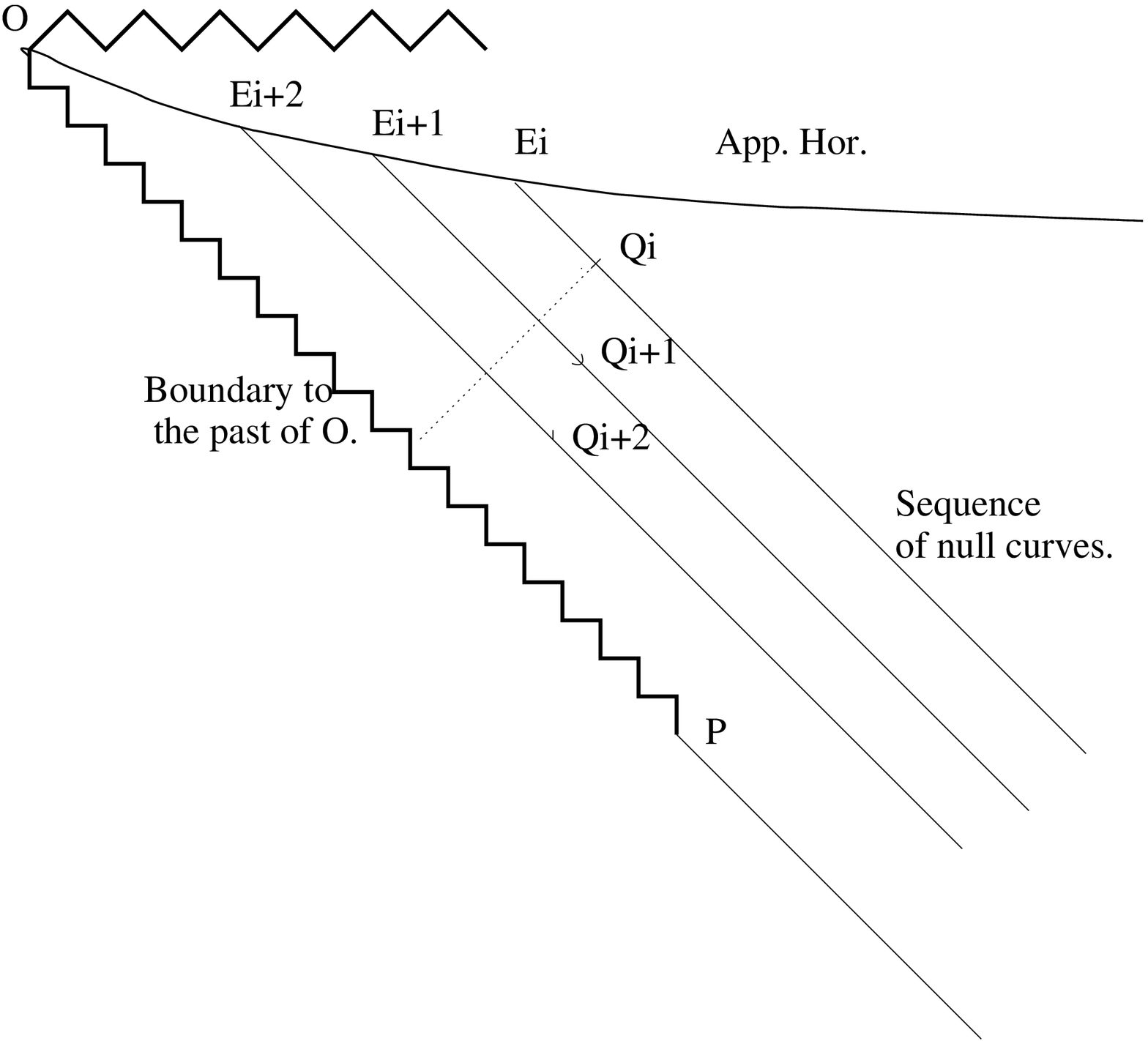}
}
\caption{Construction of a sequence of ingoing null geodesics approaching O.}
\end{figure}

In figure 3, there is no portion of the boundary beyond O that is exposed. In figure 4, however, there is a null portion.
 In this paper, we examine if O would yield the information about the
 existence of an exposed portion. 
  O is a covered point of the singular boundary (with or
 without a naked portion). This point is where the apparent horizon meets
 the singularity. Mathematically, one works with points on the apparent
 horizon in the approach to the concerned point. This could be
 looked upon as a property of the marginally trapped regions which constitute
 the apparent horizon. One is therefore working with
 outgoing  null congruences with zero expansion. If a singularity meets
 such a region, the point in the strict sense will not have any non-spacelike geodesics emerging
 from itself and will therefore be covered. 

We prove the criterion for the dust case. We offer a simple justification
 for the difference in the Penrose diagrams in  the naked and covered dust
 cases. However, the criterion is not dust specific and should be further
 examined in general cases.

 The plan of the paper  is as follows. We first motivate the view that the tangent vector to a geodesic in a congruence represents a 'flux density' analogous to the situation in ordinary electrostatics where one has lines of force.
 We wish to investigate the tangent vector on
 the apparent horizon.
 The next section describes the self similar dust model where we discuss the
 above
 and demonstrate the connection with nakedness. In the next section  we show that the results
 can be extended  to the general dust case. Finally we turn to the 
 conformal  transformation leading to the Penrose diagram for the naked case 
from spherical co-ordinates pointing out that it should diverge at the 
singular boundary.
 We argue that in a general case, this divergence would prevent ingoing 
 geodesics from reaching the point of interest which would then suggest that 
 an exposed ingoing null boundary exists.

\section {Flux of Congruence}

  The scenario of spherical dust collapse is shown in the radial
 co-rdinates in figures 1 and 2. Observe that in the naked case (figure 2) 
several
 geodesics emerge from the centre between the event and apparent 
 horizons and fall into the trapped region crossing the apparent
 horizon ( Typically this is a case of a locally naked singularity 
\footnote{ In such naked singularities, causal curves emanating from the singularity need not reach an asymptotic observer.}). 
 All along the apparent horizon in the approach to the centre, one would
 find null geodesics which would have originated at the naked centre. As drawn
 in the figure, they appear to intersect the apparent horizon in a sequence
 of points which agglomerate in the approach to the centre (in the Euclidean
 sense of the figure). If this idea of agglomeration could be made precise then
 it would be of interest to study if they (or the geodesics) tend to cluster 
 near a point
 on the apparent horizon. Such a property  would perhaps contain information
 about  the source, if existent, for the lines  of the congruence in
 a manner analogous to electrostatics. One could expect an enhancement
 of clustering of the null curves in the naked case as there is a source of
 lines in the vicinity (naked portion of the singular boundary). No such
 source at the centre exists in the covered case and the congruence could perhaps exhibit a different characteristic behaviour.
 Motivated by this, we define the quantity of flux density. 
 Given a manifold M, one defines a congruence to be a set of curves
 such that their union is M and each point of M has a single curve 
 containing it. This is analogous to the concept of lines of force in
 electrostatics. So we define a quantity analogous to electrostatic flux density for a geodesics congruence. We first affinely parametrize the congruence
 and treat the tangent vector {\bf$\xi$} like the electric field of the analogy.
  For any 3 dimensional hypersurface we define $\xi^{\mu}ds_{\mu}$ to
 be the infinitesimal flux across the infinitesimal three area $ds_{\mu}$
 of the hypersurface. The flux density is obviously the tangent vector
 itself. It is this tangent vector that we examine in the limit of
 approach to the singularity (along the apparent horizon).

It might appear from the motivation in the previous paragraph that one would attempt to check for sources of the flux ( like taking
 $\nabla {\large .} E $ in electrostatics). Instead of this, we examine it on the apparent
 horizon in an attempt to move away from the source (naked part of the singular boundary
 or set of `points' where causal curves originate on the singularity or more precisely set of ideal points to T.I.F.s \cite{penrose})
 for searching for a non-local criterion. We find that though we are unable to
 formulate any criterion away from the singularity, we are able to examine
 the point where the apparent horizon meets the boundary which is always
 (marginally) trapped and can never be a source. 

The general expression
 for tangent vector is not available in a closed analytic form. However,
 one can calculate it in a special case which we describe below and
 subsequently show how to generalize the results.
 
\section{Self Similar Dust Model}
   
   The collapse of a spherical cloud of pressureless fluid
 is given by the following metric \cite{rootspap})

 \be
 ds^{2}=dt^{2}-\frac{R'^{2}}{1+f(r)}dr^{2}-R^{2}d\Omega^{2}
\ee

 where 

$t$ and $r$ are the co-moving time and radial co-ordinates respectively.
$R(t,r)$ is called the `area radius' and a closed expression for
 this quantity which results in the dust case has enabled substantial
 progress in understanding the model.

Two free functions arise viz. $F(r) $ , called the mass function since it is
 the total mass to the interior of a shell of radius $r$ and the total
 energy function $f(r)$ which is called so because of the constraint below
which resembles a relation between kinetic and gravitational potential
 energies of a shell.
\be
\dot{R}^{2}=\frac{F}{R}+f
\ee
The source energy momentum tensor is $diag[\rho(t,r),0,0,0]$.

The solution for $R$ mentioned above is
\be
t-t_{0}(r) =-\frac{R^{3/2}}{\sqrt{F}} {\large G}\left(\frac{-Rf}{F}\right)
\ee
   
where a singularity boundary is formed at $t=t_{0}(r)$. The central
 shell focussing singularity, which is the limit as $r \rightarrow 0$
 along this locus is of interest and turns out to be naked for some initial
 data. 

 The self-similar model is the one in which $F(r)=\lambda r $ where 
 $\lambda $ is a constant (which decides if the central singularity will
 be naked or not) and $f(r)=0$.
 \footnote{ The model is referred to as self similar since there exists a 
 homothetic Killing field $t\frac{\partial}{\partial t}+ r\frac{\partial}{\partial r} $}

 We choose the scaling $t_{0}(r)=r$. A self similar co-ordinate $z=t/r$ is
 introduced. We note the expressions for $R$ and $R'$ which will be useful
 in the subsequent analysis

 \be
 R= r \lambda^{-2/3}\left(3/2\left(z-1\right)\right)^{2/3}
\ee
\be
 \label{R'eqn}R'= \left(\frac{2\lambda/3}{z-1}\right)^{1/3}\left(\frac{z-3}{2}\right)
\ee

 We cast the metric into double null co-ordinates. It is not difficult 
 to show that
\be
ds^{2}=r^{2}\left(z^{2}-R'^{2}\right)~~du~~dv
\ee
where
\be
du=\frac{dr}{r}+\frac{dz}{z-R'(z)}
\ee
\be
dv=\frac{dr}{r}+\frac{dz}{z+R'(z)}
\ee

The double null form ($ds^{2}=C^{2}(u,v)dudv$) turns out to be useful when affine  parameters
 along null geodesics are to be calculated. For instance, along an 
 outgoing radial null geodesic ($du=0$), the affine parameter is
 $\int_{u=constant}C^{2}dv$ upto a multiplicative and an additive constant.

Now let us turn to calculating the tangent vector to the outgoing null
 radial geodesic congruence, which is our primary interest.

Assume the vector to be of the form

\be
\label{xieqn}  \xi = (Q(t,r), Q(t,r)\sqrt{1+f}/R',0,0)
\ee

where $Q$ is obtained from the geodesic equation which {\bf $\xi$} has to
 satisfy. That constraint turns out to be

\be
\label{Qeqn} Q\dot{Q} + QQ' \sqrt{1+f}/R' + Q^{2}\dot{R}'/2R' = 0
\ee 

We have provided the expressions for the most general dust case here.
 One may read off the expressions for the self similar case by setting
 $f$ to zero and using equation \ref{R'eqn} for $R'$.

The equation above takes the form

\be
1/Q=\int_{u=constant}\frac{\dot{R'}}{2R'}dk ~~+~~A(u)
\ee
 where 
 $k$ is an affine parameter along outgoing radial null geodesics 
 and $u$ is the retarded null co-ordinate. $A$ is an arbitrary function
 of $u$ resulting because of the partial integration. 

 As indicated earlier,
\be
dk = r^{2}\left(z^{2}-R'^{2}\right)dv
\ee 
keeping $u$ fixed. 

Using this in equation (1/Q) we obtain

\be
1/Q= \int_{u=constant} \frac{r}{3}\frac{\left(z-3\right)^{2}}{z^{2}-1}\left[\frac{2\lambda/3}{z-1}\right]^{1/3}~~dz
\ee

 or using the fact that $du=0$ from equation (du) 

\be
1/Q= \int_{u=constant} \frac{1}{3}\frac{\left(z-3\right)^{2}}{z^{2}-1}\left[\frac{2\lambda/3}{z-1}\right]^{1/3} e^{\left[-\int\frac {db}{b-R'(b)}\right]_{z}}~~dz
\ee

The integral over $z$ is to be evaluated from $r=0$ to the apparent horizon, where we shall be interested in evaluating the tangent vector.
 The latter can be shown to be the curve $R=F$ and turns out to be the
 locus $z=1-2\lambda/3$.

The integral being over $z$, it is important to know which value of z along
 the outgoing null curve yields $r=0$, the lower limit of the integral. 
 This issue as it is shown further
 leads to the difference in the behaviour of $Q$ in the naked and covered cases. 

 Consider then the equation
\be
r=e^{\left[-\int\frac {db}{b-R'(b)}\right]_{z}}
\ee
 which we examine for $r=0$. That will happen when the integral in the
 square bracket diverges positively. Two cases can be immediately seen to
 arise. 

\underline{Case(i)} $b-R'(b)=0$ has no real root.\\
The integrand therefore does not diverge anywhere and also remains positive
 (or entirely negative) all over the real line. It can be checked that
 $b-R'(b) > 0$ for any one real $b$ which would be sufficient to claim that
 the integrand is positive. Also, $b-R'(b)$ is bounded since $b$ is to
 be limited to the nonsingular region $z<1$ ($z=1$ is the singularity 
 curve itself). So, in order that the integral diverge, the range of integration should be infinite. We have chosen to limit the final point to the apparent
 horizon $z=1-2\lambda/3$ and hence the initial point must be $z=-\infty$.

 In fact a shorter intuitive argument is possible. It is known from the
 Tolman Bondi dust model that central singularity forms at ($t_{0}(0),0$).
 If one assumes the Penrose diagram for the covered case (which is indeed
 what this case turns out to be), the null rays crossing the apparent
 horizon begin at the centre at $t<t_{0}$ which makes $z=-\infty$ there.

 \underline{Case(ii)} $b-R'(b)=0$ has atleast one real root.

 In this case, the equation implies that $r=0$ at the value of $z$ for
 which the integral in the exponent diverges. The range of integration
 for equation (1/Q second one) would then be limited at the lower end by that value of $z$. This will
 be the root (in fact the one closest to $1-2\lambda/3$, the apparent horizon).

That this is so is seen as follows.     
Equation (r) can then be re-cast using an expansion for the integrand in
 the exponent as outlined below.

Expanding R'(b) using the Taylor series about the root (called $z_{-}$)
 it can be shown that the leading order behaviour of r is as follows

 \be
r= (z-z_{-})^{\frac{1}{1-\alpha}} + O(z-z_{-})
\ee
 where 
\be
\alpha = \left(\frac{dR'(b)}{db}\right)_{b=z_{-}}
\ee

(It can be easily checked that $\alpha <0$)
   
 At $z=z_{-}$, therefore, $r$ vanishes.

Thus in conclusion of this analysis, we note that the lower limit of
 integral for $1/Q$ differs. It is $-\infty$ when $R'(b)-b=0 $ can never have
 a real solution and is the root (closest to apparent horizon) when a 
solution exists. 

This observation plays the key role in further analysis. Making note of this
 consider equation (1/Q second one). Analyzing the various factors
 in the integrand one finds that the integrand would diverge if $z=-1$
 ( $z \ne 1 $ since we are not on the singular boundary). 
 
 $z$ will take the value $-1$  in case i. In case ii,
 the following takes place. Consider $b-R'$ using equation (R'). It is easy
 to see that $ b-R' < 0$ for all $b<0$. So, the root $b_{-}$ cannot be
 negative. Hence it is certainly greater than $-1$. Thus $z$ cannot take the
 value $-1$ in case ii in the integral for $1/Q$.

Thus the integrand diverges as $1/(z+1)$ in case i and is finite in case ii.

Expanding the rest of the integrand factor in a Taylor series about $z=-1$,
 one can easily check that integral diverges logarithmically in case i
 while staying finite in case ii.

Thus, $Q$ vanishes in case ii and stays non zero (and finite) in case ii.

    Returning now to equation \ref{xieqn} ,  we can now see that $\xi$ behaves 
 in different ways in case i and case ii on the apparent horizon, in particular
 as  one approaches the point O on the Penrose diagrams shown (figures 3 and 4
\footnote{Figure 4 is a diagram of a locally naked singularity. The self similar cloud which we examine here turns out to be globally naked. However, the
 structure near O is the same as any locally naked case and figure 4 can be used.})
. It can be checked that the factor $\sqrt{1+f}/R'$ tends to is a non-zero
 finite quantity on the apparent horizon.

  So, {\bf $\xi$} vanishes in case i and tends to  a non-zero quantity
  in  case ii. 

  Using its interpretation of flux density of the congruence, we find that
 the congruence tends to cluster in case ii as against case i. 

  From previous analysis of naked singularities (self similar cases) using
  analysis for emergence of  geodesics (roots analysis), it can be checked
 that case i corresponds  to the covered case and case  ii  corresponds to
  the naked  singular  metric.

 \section{ Extension to the general dust case}

 In the general  dust case,  the equation \ref{Qeqn} yields no closed analytic
 solution which would have clearly been  useful. However, we note that
 we  are interested only  in the  behaviour of $Q$ in the limit
 of approach to point O on the apparent horizon.

To this end the following observation plays an important role. It is
 shown that given a dust solution, one can construct a modified dust
 solution (modified distribution) which in a  suitable limit approaches
 the given dust solution \cite{divpap}.  The key result that makes
 this construction useful is that it is proved that naked  modified
 distributions reproduce naked dust solutions given and  covered modified
 distributions reproduce covered ones. One can then work with the
 modified distribution for  the  given dust solution and take the limit
 which preserves naked or covered nature. We outline the construction
 in \cite{divpap} below

 a) Marginally bound case  ($f =0$)

   Imagine a shell of radius $r_c$ in  the given  Tolman Bondi dust
 model. Replace the interior of the shell by a  self similar dust metric,
 matching the first and second fundamental forms at the interface  $r=r_c$.
 It can be shown  that  this restricts the self similarity parameter 
 $\lambda$ which  appears in the mass function.  This specifies the
 self similar solution  completely. Now taking the limit  as $r_c$ tends to
 zero, one can  show \cite{divpap} that the  matching constraint does
 imply that the interior self similar solution stays naked in the limit
 if the original dust solution was naked and like wise in the covered case.

b) Non Marginally bound case ($f\ne 0$)

    The construction is similar in this  case except for  an additional  
 interface.  Two shells, $r_c1$ and $r_c2$ ( say  $r_c1 < r_c2$ )  are  now
 considered. To the interior of $r_c1$,  we replace by a self similar metric.
 Between  the two shells, we replace  with  a dust portion having  $f$ so
 behaved that it increases smoothly from zero at $r_c1$ to $f(r_c2)$ of the 
 original dust metric. The $F$ function for this extra portion of
 dust however  is  the same as that of the original dust metric.
 We match the first and second fundamental forms
 at each of the interfaces. As  before, this can be shown to  constrain 
 the interior self similar solution uniquely given $r_c1$ and the original
 dust solution. Again, the property of being naked or covered is preserved
 in the limit ( $r_c2 \rightarrow 0$) like the previous case  \cite{divpap}.

We now consider $Q$ in the modified distribution for  any  given dust
  solution. In  the self similar  part  of the latter, results of the
 previous section  apply. Since the congruence of  outgoing geodesics is
 smooth, so  is $Q$. This makes  $Q$ continuous across the interface/s
 in the modified distribution. Now imagine the given dust solution as
 the  limiting case of the modified distribution. In the limit of approach
  to point  O on the apparent horizon, one  has to evaluate $Q$ in the
 self similar part.  Because of continuity of $Q$, the same  behaviour
 will continue to hold  in the limit of the interface/s tending to zero  when 
  the original dust  solution is reproduced. Making use of the fact that
 the property  of being naked or covered is preserved in this limit, one
 concludes that the behaviour of $Q$ in the self similar naked  and covered
  cases continues to hold in the general dust scenario  as  well.

\section {Conformal transformation and Penrose diagram}

 The tendency of the null geodesics of the congruence to cluster
  in  the  approach to the singular  boundary is basically due to
 the inappropriate  nature of the co-ordinate  system at the boundary.
 If one wishes  to depict the boundary as a curve in a particular
  co-ordinate system, the  null congruence has to be well defined
 ( in the sense  that the property that one and only one curve passes
  through every point should hold  even when the congruence is extended
  to the boundary). For instance, in the naked dust case,  when  one  uses
  spherical co-ordinates it can be seen  that  several radial null  geodesics
  appear to emerge from the central singularity with the same tangent
   vector \cite{simple}.
 
  The  issue about the co-ordinate system  being appropriate for such an 
 extension  could thus related to the behaviour of {\bf $\xi$}. 

  From a technical point of view, the calculations using the radial
 co-ordinates could  be performed  in a conformally related metric which
 avoids the problem of clustering if it occurs. The conformal  transformation
 would be the  one leading to the structure  of the singularity as  depicted 
 in the Penrose diagram. 

 We  now argue that the under a conformal transformation which diverges
 in the limit of O, {\bf $\xi$} which tend to a non-zero limit transform
 to vector fields which  vanish in the  limit. 

 Recall  that we defined {\bf $\xi$} for any geodesic  congruence using
 an affine parametrization. Under conformal  transformations,  affine parameters along null geodesics change ( unlike timelike geodesics which do  not
 remain geodesic curves, null  geodesics do stay so provided  the affine parameter changes appropriately). Infinitesimal parameter $ds$ transforms to
 $\Omega^{2}(x^{\mu})ds$ \cite{waldgr}, where $\Omega^{2}$ is a conformal transformation. Thus it is obvious that 
 $\xi^{\mu}=dx^{\mu}/ds$ if finite and  non-vanishing in the limit will
 vanish under $\Omega^{2}$ transformation provided the  latter diverges there.
  
 Thus we find that at least in the dust case, one requires a conformal
 transformation which diverges on the apparent horizon in the  limit of approach
 to the singularity in the naked case as  against the covered case where
  the radial co-ordinates are appropriate to describe the singularity
 structure.\footnote{ It is evident from  the definitions that the difference 
in the behaviour of 
 flux is due to
 difference in the metrics, one being conformally related to the other.
 This raises the question of the flux being an appropriate characterization
 of the causal structure. It certainly is not appropriate at any event 
 within the spacetime but in the limit of approach of the boundary
 its behaviour indicates if a metric with the correct limiting causal
 structure has been employed in its calculation.} This justifies the difference in the structure of the singular boundary near O in figures 3 and 4.

\section {A possible general scenario}

 Consider the cases of collapse  in which the singularity formed meets the
 boundary of the trapped region ( or even crossing it  as in naked cases)
 Now  it  would be  of interest to examine {\bf $\xi$} in general on the apparent
 horizon and check if  it  vanishes or not
 in the approach to  the singular boundary. If it does not,  then one
 invokes the diverging conformal  transformation to  obtain the
 correct causal depiction.
  The immediate question would be the naked  or
 covered nature of such a singularity.  We certainly  know  that it is naked
  in the dust case when the conformal  transformation diverges. We present
 an argument suggesting its validity in a general  scenario relaxing the assumption of 
 dust and spherical symmetry. \\

   Let point O be the intersection of the singularity and the apparent horizon
 as before.\\ \\
  {\it Theorem}: No ingoing  null geodesic can reach point O after a conformal transformation if  the conformal
          transformation diverges  at O. \\ \\
 {\it Proof}:\\ 
   Consider a space-like hypersurface from which an ingoing null geodesic
 reaches O if possible.
   From the Raychaudhari equations , it can be shown that once a  null geodesic
 has negative expansion,  it will reach a conjugate point 
 after a finite  amount of affine parameter  has elapsed. If  the conformal
 transformation  diverges in  the limit, then a null geodesic reaching O
  would imply an elapse of infinite amount of affine parameter\footnote{See previous section}. This is
 a contradiction. Hence the conjugate point must occur before O on the null
 curve, beyond  which the geodesic cannot be extended. So the geodesic cannot
 reach O. $\Box$ \\ \\  

{\it Lemma}: There exists an ingoing null boundary (including O) to the past 
 of O  if no ingoing geodesic reaches O. \\ \\
{\it Proof}:  \\
     Consider a sequence of ingoing null geodesic segments $\{\Lambda_{n}\}$ 
 with future endpoints on
 the apparent horizon, the endpoints approaching O as $n \rightarrow \infty$. 
 Let there be no ingoing null boundary to the past of O, if possible. Then
 there will be a limiting null geodesic of $\{\Lambda_{n}\}$ which reaches O.
This contradicts the previous theorem. $\Box$\\ \\
  In the Penrose diagram (figure 4) one can imagine an ingoing null geodesic
 which reaches the null singular boundary at point P. This is the conjugate
 point for that geodesic. We have simply justified that there 
 will be a boundary to  the spacetime in place of the geodesic curve between 
 P and O.   \\
If the apparent horizon is spacelike  in the approach
 to  O, the above portion of boundary is certainly exposed  into the untrapped
  region and is therefore naked. One may ask if the apparent horizon
  is always spacelike in the naked case. (It is certainly true for dust 
\cite{jhingan}). If a singular portion to the past of O exists, then it cannot
 be trapped. If such a portion existed then it would simply appear from a
 Penrose diagram with such a portion that causal curves
 would emanate from them. These arguments are made precise below (For 
 definitions of IFs, Proper IFs, and TIFs, see \cite{penrose},\cite{haw}).

 {\it Theorem}: Existence of an ingoing null boundary to the
 past of O implies the existence of TIFs. ( The portion of the boundary
 contains the ideal point of a TIF)\\ \\
  {\it Proof}: \\
    Consider a sequence $\{\Lambda_{n}\}$ as before. This choice is fixed 
 throughout this proof. Also note  that $J^{+}(\Lambda_{i}) \subset J^{+}(\Lambda_{j})$ for all $j>i$ (The null curves are successively to the past). Since each $\Lambda_{i}$
 has a future endpoint on the apparent horizon $E_{i}$, all points 
 of $\Lambda_{i}$ except $E_{i}$ are untrapped. Choose one such point $Q_{i}$.
 $I^{+}(Q_{i})$
 is non empty. By choosing $Q_{i+1}$ to the causal past of $Q_{i}$ for every 
$i$ ( we can always do that since the null curves are successively to the past), one
 obtains a sequence of (proper) indecomposable future sets of $Q_{i}$ which 
 are nested. (One may begin the sequence at any $i$)
 The limiting IF as $ i \rightarrow \infty$ is therefore non empty.\\ This
 IF will be proper iff there is a limit of $\{Q_{n}\}$ which is a part of
 spacetime. There is a boundary to the past of O. Therefore at least one Q
 sequence (constructed as described above) exists which fails to have a 
limiting Q within spacetime. 
\footnote{
 Observe that if $\{Q_{n}\}$ are constructed as defined above, then
 for every neighbourhood of each $E_{i}$, one can find a sequence $\{Q_{n}\}$
 such that $Q_{i}$ lies in that neighbourhood (to the past of $E_{i}$) simply
 by beginning the sequence at that $i$. This
 is true for every $i$. Next, the union of limit points of all possible Q sequences
 defines the `first' ingoing null geodesic to reach the singularity ( ingoing
 null curve at P in figure 6). 
 If the limit for every Q sequence exists in spacetime,
 then
 one obtains limit points within spacetime as close to O as possible. However, there is
 a null boundary to the past of O and so this is not possible.} Consider
 this Q sequence ( there are actually an uncountably infinite of them). The corresponding limiting IF of the $Q_{i}$s will be
 a TIF since there is no point in spacetime of which it is the future.  $\Box$

\section{Summary and Conclusion}

 The tangent vector field to a null geodesic congruence being thought of as 
`flux density' of a congruence of geodesics is 
 examined for behaviour on the apparent horizon in the approach to
 the singularity (point O) in the dust collapse model. There is a correlation
 with the property of nakedness with this behaviour. Demanding that the
 vector vanishes at the covered point O forces the divergence of the conformal
 transformation at O which leads to the Penrose diagram for the naked scenario.
 Since the flux vanishes in the covered case, there is no such divergence
 and hence the Penrose diagrams in the two cases differ.
        
 One demands that the flux vanishes at O in a general collapse scenario on the grounds that O is covered when calculated using a metric exhibiting the correct causal structure,
 and in case it does not, one uses a suitable conformal transformation (i.e
 one which diverges at O) in order to obtain the correct causal structure
 near O. We show that there will be no ingoing null geodesic reaching O
 if the latter is the case and argue that it indicates the existence of
 a portion of singularity which is untrapped.

 In conclusion, we have shown that the information about whether the singularity formed in collapse is naked is contained at the intersection of the apparent
 horizon and singular boundary in the spherical dust case.
 We also suggest that it holds in the case of a general
 collapse. It should also be noted that the procedure of checking if an
 appropriate conformal transformation is necessary does not directly involve 
 checking
 for emergence of causal curves from the singularity.

 \end{document}